\documentclass[twocolumn,snoshowpacs,pre,preprintnumbers,superscriptaddress,amsmath,amssymb,floatfix]
{revtex4}
\usepackage{graphicx}
\usepackage[caption=false]{subfig}
\usepackage{color}
\usepackage{placeins}
\usepackage{tikz}
\usepackage{mathtools}
\usepackage{amsthm}
\usepackage{romannum}
\usepackage{color}
\usepackage{xkeyval,xcolor}
\usepackage{amsmath}
\usepackage{caption}
\usepackage{epstopdf}

\begin{document}

\title{Optimization of robustness based on reinforced nodes in a modular network}

\author{Yael Kfir-Cohen}
\affiliation{Bar-Ilan University, Ramat Gan, Israel}
\author{Dana Vaknin}
\affiliation{Bar-Ilan University, Ramat Gan, Israel}
\author{Shlomo Havlin} 
\affiliation{Bar-Ilan University, Ramat Gan, Israel}
\date{\today}

\begin{abstract}
    Many systems such as critical infrastructure exhibit a modular structure with many links within the modules and few links between them. 
    One approach to increase the robustness of these systems is to reinforce a fraction of the nodes in each module, so that the reinforced nodes provide additional needed sources for themselves as well as for their nearby neighborhood.
    Since reinforcing a node can be an expensive task, the efficiency of the decentralization process by reinforced nodes is vital.
	In our study we analyze a new model which combines both above mentioned features of real complex systems - modularity and reinforced nodes. 
	Using tools from percolation theory, we derived an analytical solution for any partition of reinforced nodes; between nodes which have links that connect them to other modules (``inter-nodes") and nodes which have connections only within their modules (``intra-nodes"). 
	Among our results, we find that near the critical percolation point ($p\approx p_c$) the robustness is greatly affected by the distribution. In particular, we find a partition of reinforced nodes which yields an optimal robustness and we show that the optimal partition remains constant for high average degrees.
\end{abstract}
  \maketitle

\section{INTRODUCTION}

In recent years, much attention has been focused on the resilience and stability of networks with a community structure \cite{leichtdsouza2009,wang-pre2013, radicchi-naturephysics2013,shai2014resilience,Shai2015,Shekhtman_2015}.
The resilience of a network can be estimated by the size of the connected giant component (GC) after failures, where the GC is called the order parameter of the system in percolation theory \cite{Coniglio_1977,coniglio-jphysicsa1982,Sokolov_1986,staufferaharony,bunde1991fractals,morone-nature2015,mureddu-scientificreports2016}. 
This is based on the assumption that when nodes are not connected to the GC they are not regarded as functioning, since they can not communicate or get resources from other nodes. This condition that functioning depends on being connected to the GC can be regarded as a centralization feature. Examples of real-world networks with a community structure are the brain \cite{Meunier2010,Stam2010,Morone2017}, infrastructures \cite{eriksen2003modularity,Guimera2005} and social networks \cite{Girvan2002,Thiemann2010,Liu2012} as well as many others \cite{Onnela2007,Gonzlez2007,Lancichinetti_2009,Mucha2010,barthelemy-physicsreports2011}.

\begin{figure}[ht!]
\centering
\includegraphics[scale=0.49]{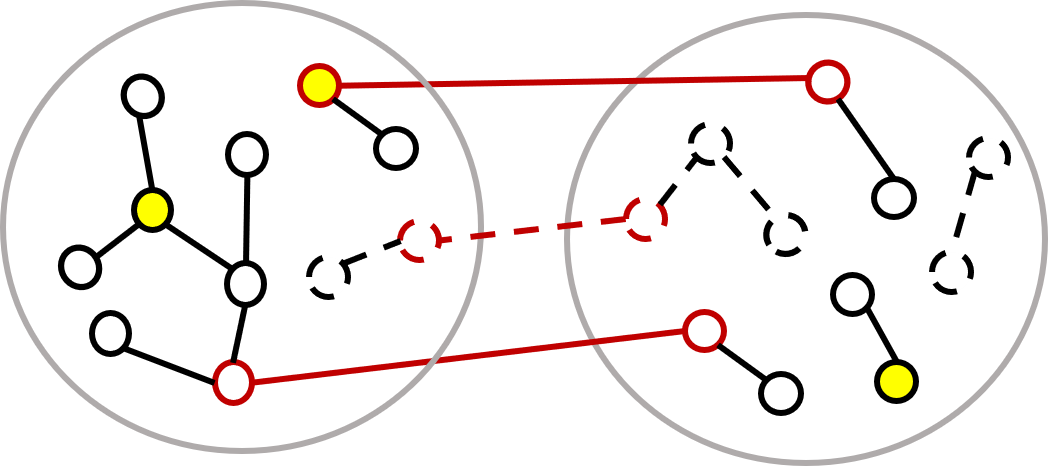}
\caption{ A schematic representation of the model with m=2 modules. Each grey circle represents a module. The red circles are inter-connected nodes, which have red links to the inter-connected nodes of the other module. Inside each module, the nodes are connected through black intra-links. The yellow nodes are reinforced nodes while some are also inter-connected nodes and some are not inter-connected nodes.
The functional component contains the giant cluster and the clusters which contain reinforced (yellow) nodes. Nodes and links which are not part of the functional component are marked by dashed lines.}
\label{fig:model}
\end{figure}

Here we focus on a recently proposed community model \cite{Dong2018} where only a fraction $r$ of nodes are capable of having inter-links that connect them to other modules (communities).
In addition, a new concept of reinforced nodes has been recently introduced into modeling of real-world networks \cite{Yuan2017}. 
The reinforced nodes are nodes that have their own support and can also support the cluster of nodes connected to them. 
Therefore, even if they are disconnected from the giant component, they are still regarded as functional.
For example, in the internet network, communication satellites \cite{Henderson1999} or high-altitude platforms \cite{Mohammed2011} can serve as reinforced nodes and support important internet ports in cases of connection failures. Thus, the concept of reinforced nodes actually relaxes the centralization feature of the network, since there are nodes that are not in the GC but can still function properly. 
Thus the process of reinforced nodes can be regarded as a decentralization feature.
Interestingly, it has been found \cite{Yuan2017} that in a regular, non modular network a very small fraction of reinforced nodes increases the robustness significantly. 
Considering reinforced nodes, the new order parameter which expresses the functionality of the system can be taken as the $functional$ component and not the known $giant$ component \cite{Yuan2017}.
The functional component contains both the giant component and smaller components which include at least one reinforced node, see Fig. \ref{fig:model}.

Here, we study the stability of a modular system in the presence of reinforced nodes, where the stability is characterized by the size of the functional component.
In particular, we distinguish between reinforced nodes that are connected to other modules (inter-connected nodes) and reinforced nodes that are connected only to their own modules (intra-connected nodes). 
We find the functional component size using both simulations and theory, and we address the following optimization question: where to place the reinforced nodes in order to optimize the robustness of the system to random failures (i.e., to obtain the largest functional component)?
Specifically, the question is how to distribute the reinforced nodes between intra-connected nodes and inter-connected nodes?
An example of a system that motivates our research is the network of power grids which have electric generators. 
In this example, cities can be regarded as communities and the reinforced nodes are the electric generators.
Thus, the optimization question is how to distribute the electric generators between the inter-connected power poles that connect different cities and the intra-connected power poles that do not.

\section{Model and theoretical approach}

Our network model, of size $N$, consists of $m$ ER modules (communities) where each module has $N/m$ nodes and the average intra-degree for a node is $z$.
We randomly select a fraction $r$ of nodes in each module to become inter-connected nodes, and place $M_{inter}$ inter-connected links between any two modules. 
For each pair of modules, $A$ and $B$ for instance, each inter-connected link is randomly placed between an inter-connected node from $A$ and an inter-connected node from $B$. Thus, the inter-degree distributions are Poisson with average inter-degree $\kappa = {m\*M_{inter}}/{rN}$.

In addition, we use the following notations: $\rho$ is the fraction of reinforced nodes in the network, $\rho_x$ is the fraction of reinforced nodes in the network that are inter-connected nodes and $\rho_o$ is the fraction of reinforced nodes in the network that are intra-connected nodes, thus, $\rho_x+\rho_0=\rho$. 

Next, we derive the size of the functional component (FC) under attack as a function of the number of reinforced nodes, first for a random distribution of reinforced nodes and later for a particular distribution of them.
We use tools of percolation theory and define the generating functions for the intra- and inter- nodes as follows: $G^{intra}_{0}$ and $G^{inter}_{0}$ for the degree distribution and $G^{intra}_{1}$ and $G^{inter}_{1}$ for the excess degree distribution \cite{Gross2020}.
We define $u$ as the probability that an intra-link leads to a node that is not part of the FC; $v$ as the probability that an inter-connected link leads to a node that is not part of the FC, and $S$ as the probability that a node is part of the FC.
Thus, for a network where the fraction of reinforced nodes, $\rho$, is distributed randomly, the probabilities $u$, $v$ and $S$ satisfy the equations:

\begin{equation}
\label{eq:uvs}
\begin{gathered}
1-u=p\left[1-(1-\rho)G^{intra}_{1}(u)[1-r+r\prod_{}^{m-1} G^{inter}_{0}(v)]\right]\\
1-v=p\left[1-(1-\rho)G^{intra}_{0}(u)\prod_{}^{m-1} G^{inter}_{1}(v)\right]\\
S=p\left[1-(1-\rho)G^{intra}_{0}(u)[1-r+r\prod_{}^{m-1} G^{inter}_{0}(v)]\right],
\end{gathered}
\end{equation}
for the case of randomly removing a fraction $1-p$ of the nodes from the network.

In our model, the inter-links are connected randomly, therefore the distributions are Poissonians with average inter-degree $\kappa$, i.e. $G^{inter}_{0}(x) = G^{inter}_{1}(x) = e^{-\kappa(1-x)}$. 

In addition, in the limit of infinitely large ER networks the degree distributions for the intra-connected nodes are Poissonians with an average intra-degree $z$ and thus $G^{intra}_{0}(x) = G^{intra}_{1}(x) = e^{-z(1-x)}$. 
These equations (for the random case) lead to a single transcendental equation relating $S$, $r$ and $\rho$:
\begin{equation}
\label{eq:communities_reinforced}
\begin{multlined}
e^{-zS}(r-1)(1-\rho)+1-\frac{S}{p}=r(1-\rho)e^{-zS}\cdot \\
\exp\left[\frac{(m-1)p\kappa\left(e^{-zS}(r-1)(1-\rho)+1-\frac{S}{p}-r\right)}{r}\right].
\end{multlined}
\end{equation} 
Note that this equation is a generalization of Ref. \cite{Yuan2017}, where the network contains only a single ER module ($m = 1$), to the case of any number of $m$ modules.
In addition, this equation is a generalization to the case of $\rho$ reinforced nodes of Ref. \cite{Dong2018} which analyses a modular network without reinforced nodes, i.e., $\rho = 0$.

For any partition of the fraction of reinforced nodes $\rho$ between the intra- and inter-nodes ($\rho_o$ and $\rho_x$ respectively), $u$, $v$ and $S$ fulfill the following equations, 

\begin{equation}
\label{eq:uvs_rho_x}
\begin{gathered}
\hspace*{-0.3cm}
1-u=p\left[1-G^{intra}_{1}(u)[1-r-\rho_o+(r-\rho_x)\prod_{}^{m-1}G^{inter}_{0}(v)]\right]\\
1-v=p\left[1-(1-\frac{\rho_x}{r})G^{intra}_{0}(u)\prod_{}^{m-1} G^{inter}_{1}(v) \right]\\
\hspace*{-0.3cm}
S=p\left[1-G^{intra}_{0}(u)[1-r-\rho_o+(r-\rho_x)\prod_{}^{m-1} G^{inter}_{0}(v)]\right].
\end{gathered}
\end{equation}

The generating functions for the intra- and inter-nodes are the ones defined above. Therefore, a single transcendental equation for the functional component relating $S$, $r$, $\rho_x$ and $\rho_o$ can be written as:

\begin{equation}
\label{eq:m_communities_rho_x_rho_o}
\begin{multlined}
e^{-zS}(r-1+\rho_o)+1-\frac{S}{p} = (r-\rho_x)e^{-zS}\cdot \\ 
 \exp \left[\frac{(m-1)p\kappa\left(e^{-zS}(r-1+\rho_o)+1-\frac{S}{p}-r\right)}{r}\right]. 
\end{multlined}
\end{equation}

\section{RESULTS}

\begin{figure}[ht!]
\centering
\includegraphics[width=1\linewidth]{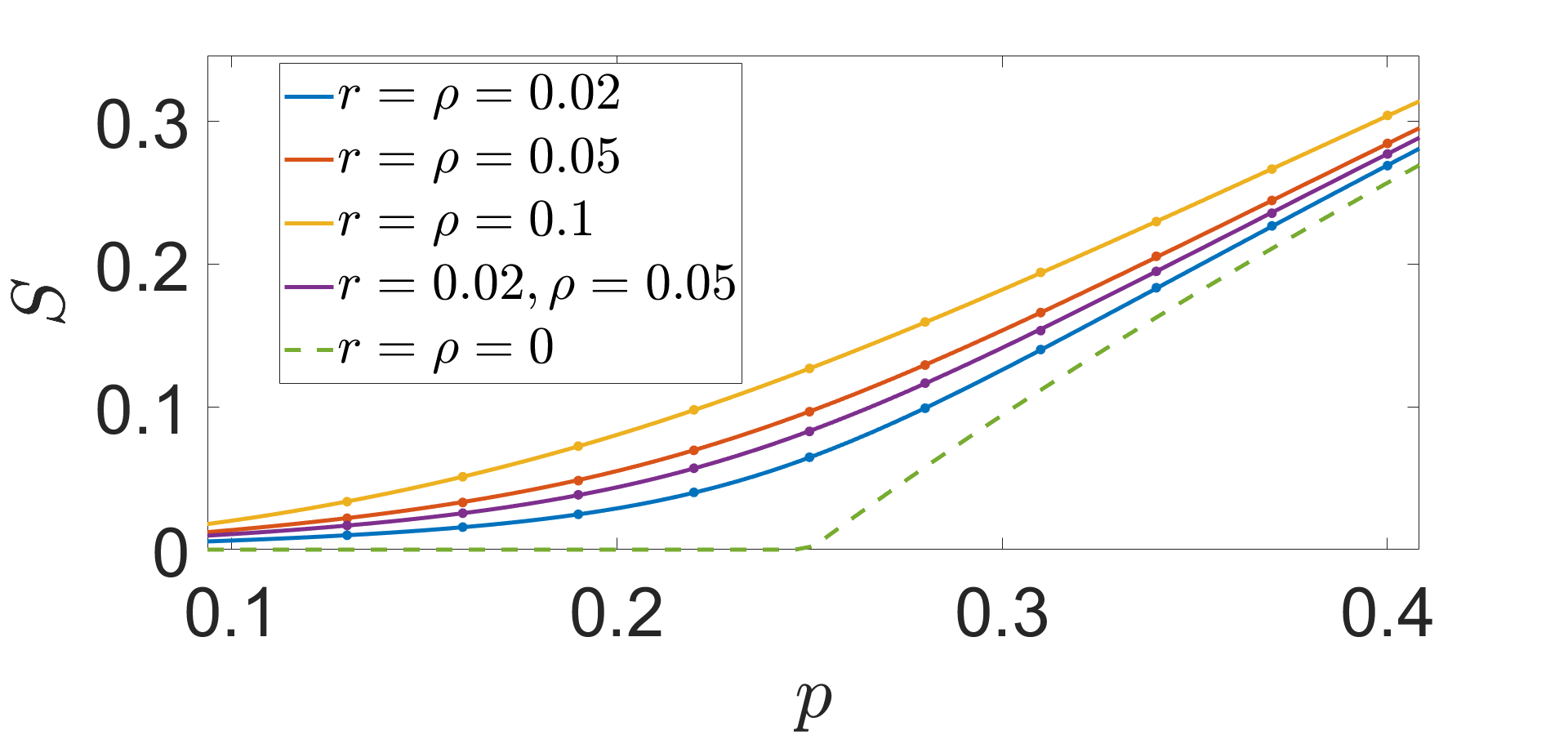}
\caption{The size of the functional component, $S$, as a function of $p$ for different $r$ and $\rho$ values where the reinforced nodes are distributed randomly.
Lines and symbols denote analytical and simulation results, respectively.
The dashed green line represents the theoretical solution of a single ER network without reinforced nodes. 
For these runs we chose, $m=2$, $N_1=N_2 = 10^6$, $M_{inter} = N_1$ and $z=4$.}
\label{fig:both}
\end{figure}

\begin{figure}[ht!]
\centering
\includegraphics[width=1\linewidth]{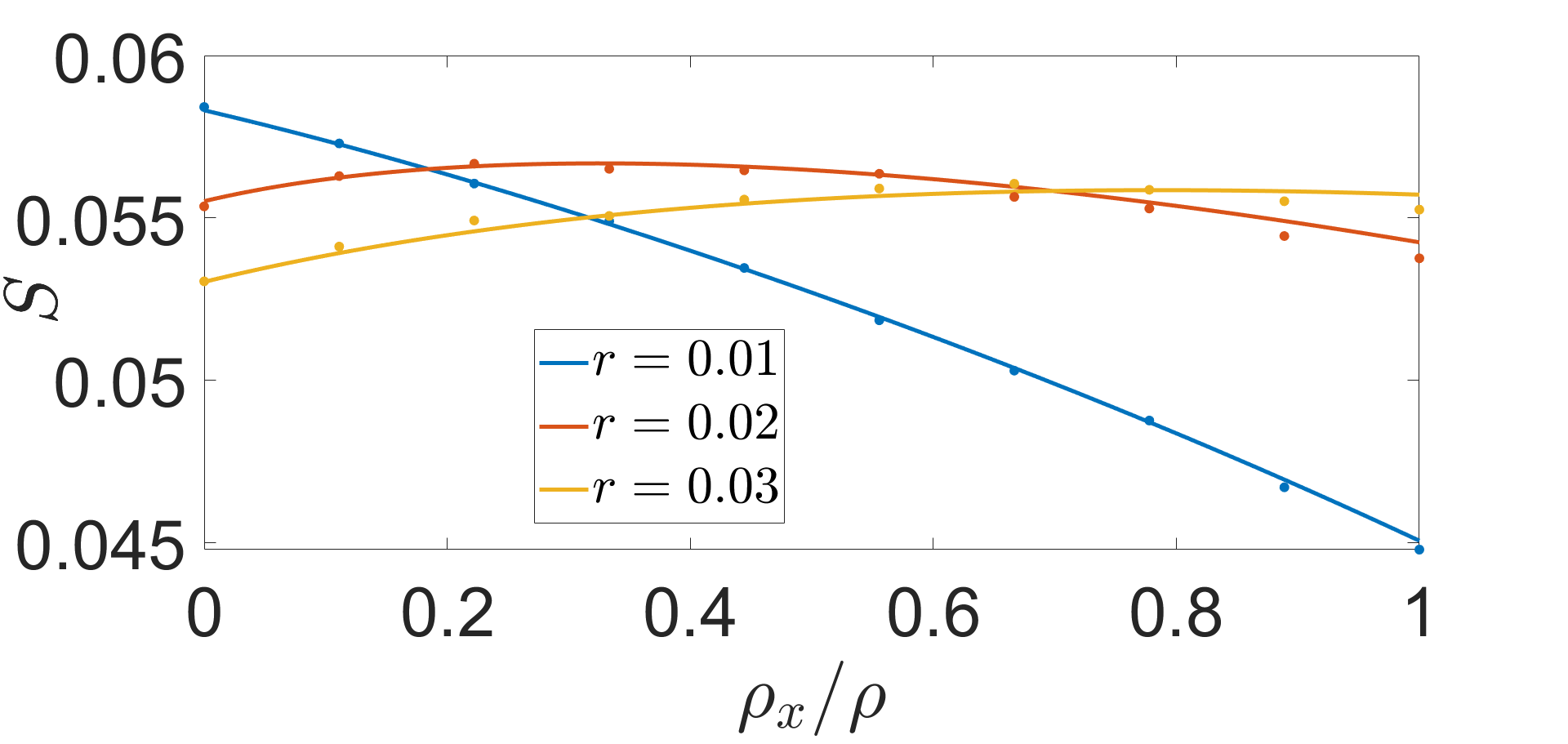}
\caption{The size of the functional component, $S$, as a function of $\rho_x/\rho$, at $p_c$, for several values of $r$. Lines and symbols denote analytical and simulation results, respectively. 
Here, $m=2$, $N_1 = N_2 = 10^6$, $M_{inter} = 5\cdot 10^4$, $z=3$, $p_c=0.3333$ and $\rho = 0.01$.}
	\label{fig:S vs rho_x at pc}
\end{figure}

\begin{figure}[ht!]
\centering
	\subfloat[]{\includegraphics[width=\linewidth]{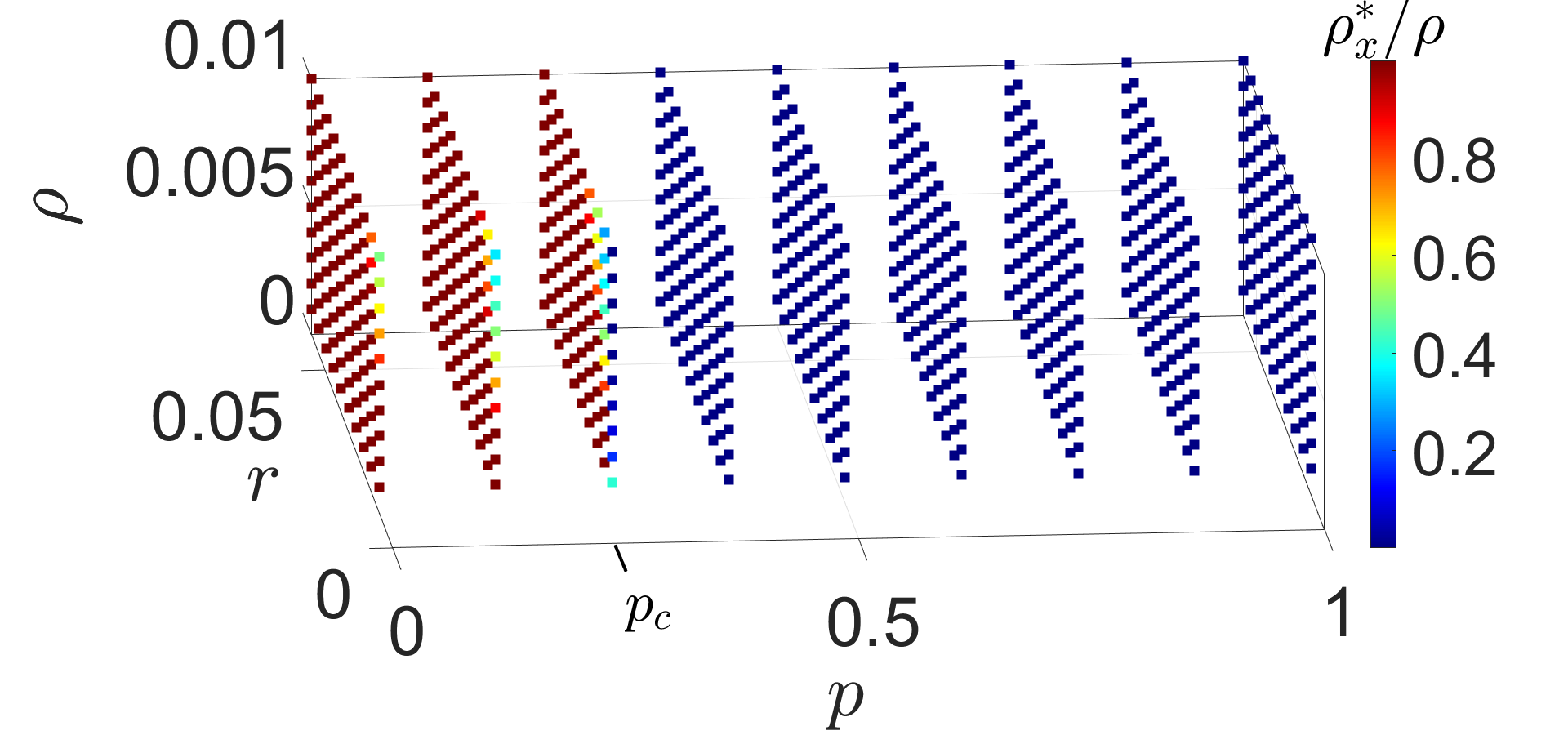}}\\
	\subfloat[]{\includegraphics[width=\linewidth]{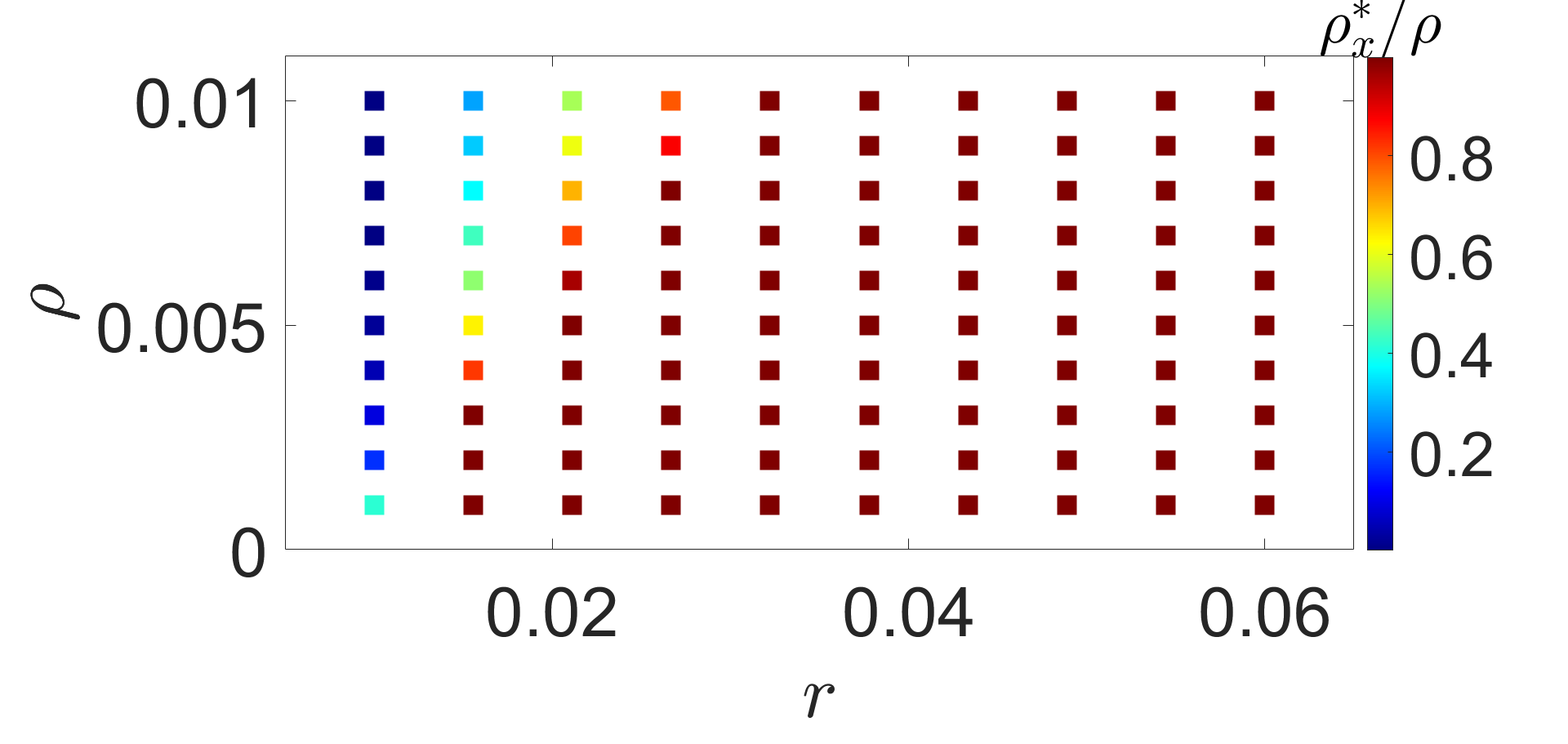}}\\
\caption{The value of $\rho^{*}_x/\rho$ as a function of $r$ and $\rho$, for $m=2$, $N_1 = N_2 = 10^6$, $M_{inter} = 5\cdot 10^4$, $z=4$.
    {\bf (a)} $\rho^{*}_x/\rho$ for different values of $p$, and {\bf (b)} $\rho^{*}_x/\rho$ only for $p=p_c=\frac{1}{z}$.}
	\label{fig:best_rhox}	
\end{figure}

\begin{figure}[ht!]
\centering
\includegraphics[width=\linewidth]{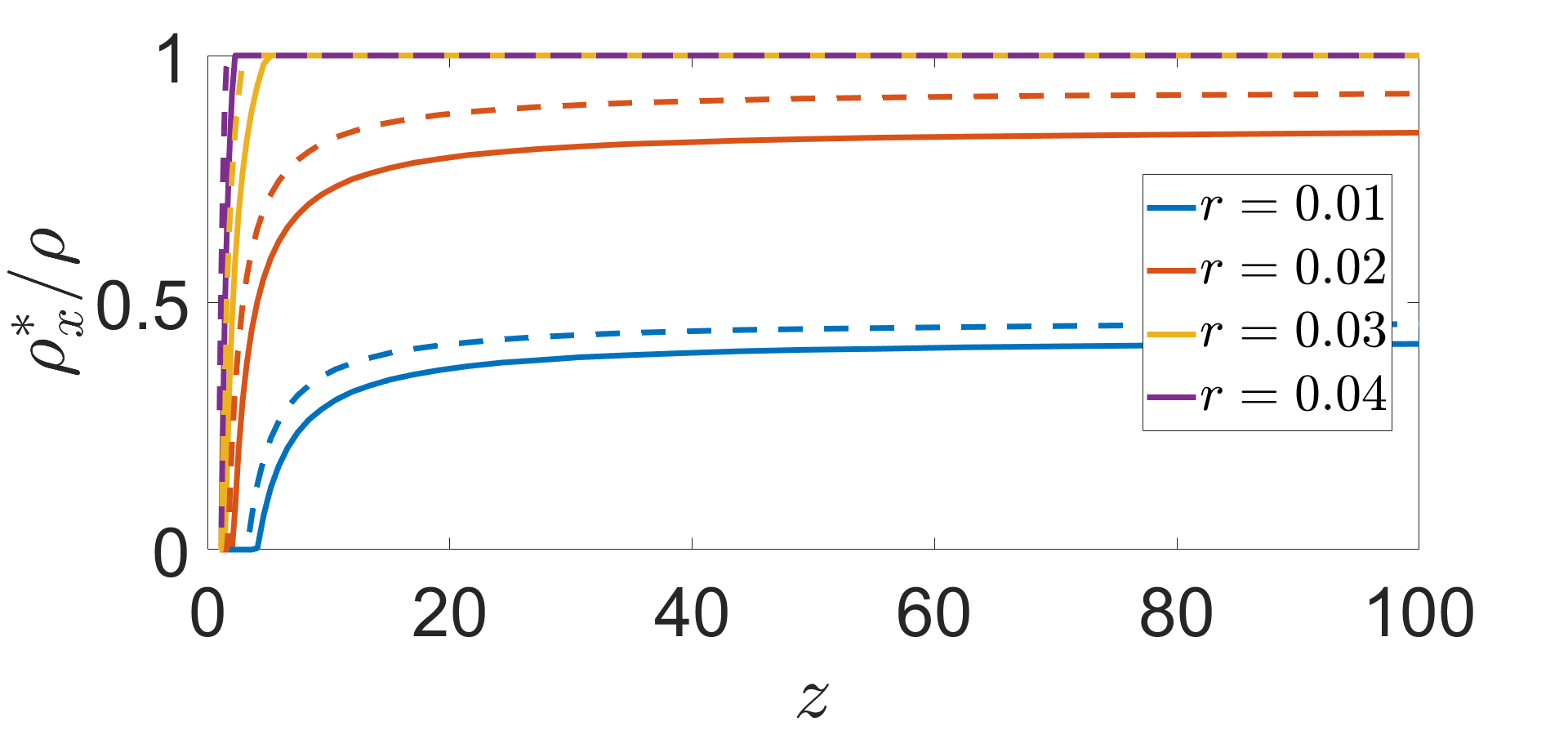}
\caption{The fraction $\rho^{*}_x$/$\rho$ as a function of $z$ for different values of $r$. Here, $m=2$, $N_1 = N_2 = 10^6$, $M_{inter} = 5\cdot 10^4$ and $\rho = 0.01$. Note that the optimal ratio of the reinforced nodes approaches constant for high intra-degrees. The dashed lines are for $p=0.9\cdot p_c$ while the full lines are for $p=p_c$.}
\label{fig:best_rhox_k}
\end{figure}

For the sake of simplicity, here we analyze a network with $m=2$ modules, while in the SI (B) we refer to the general case of $m$ modules.
We quantify the resilience of a network where the reinforced nodes are positioned randomly, by obtaining the size of the functional component $S$ for different values of $p$, $r$ and $\rho$, both by solving Eq. \ref{eq:communities_reinforced} and by numerical simulations. 
In Fig. \ref{fig:both}, we present $S$ as a function of $p$, and show that $S$ increases both with the increase in $r$ and the increase in $\rho$.
As seen, the analytical solution is in a very good agreement with the results obtained from the numerical simulations.
Next, we study various networks with different distributions of the reinforced nodes between inter- and intra-nodes. 
In Fig. \ref{fig:S vs rho_x at pc}, we show for $p=p_c$ the functional component, $S$, as a function of $\rho_x/\rho$ (i.e. the fraction of reinforced nodes which are also inter-connected nodes). 
It can be seen that for a given $\rho$ and an average intra-degree $z$, $S$ as a function of $\rho_x/\rho$ behaves differently for different values of $r$.
For a very small $r$, $S$ decreases monotonically; for a slightly larger $r$, $S$ increases monotonically. However, for intermediate values of $r$, $S$ behaves as a concave function with a maximum.
Thus, we conclude that for very small $r$ values the best strategy is to place the reinforced nodes as the intra-connected nodes, while for slightly larger $r$ values it is better to place them as the inter-connected nodes.

Then, for any given $\rho$ we find its partition of $\rho_x$ and $\rho_o$ which generates the maximal functional component.
We define $\rho_x^*$ as the value of $\rho_x$ which yields the optimal division i.e., we calculate the FC, $S$, by Eq. \ref{eq:m_communities_rho_x_rho_o} for different values of $\rho_x$ between $0$ to $\rho$ and define $\rho_x^*$ to be the $\rho_x$ value which maintains the maximal $S$ value.
We calculate $\rho_x^*$ for different values of $r$, $\rho$ and $p$ (see Fig. \ref{fig:best_rhox}(a)).
For $p>p_c$, we obtain $\rho_x^*=0$ (i.e., it is better to reinforce the intra-nodes) for any values of $r$ and $\rho$. 
On the other hand, for $p\leq p_c$, $\rho_x^*$ is determined by $r$ and $\rho$. For any given $\rho$, $\rho_x^*$ increases with $r$, see for instance Fig. \ref{fig:best_rhox}(b) for $p=p_c$. Our results in Fig. \ref{fig:best_rhox} demonstrate that one can distribute the reinforced nodes between the intra- and inter-nodes such that the robustness is optimal.
In Appendix A we show that the differences in the size of the FC between different divisions of the reinforced nodes are mostly significant for $p\leq p_c$ values. Thus, when demonstrating the optimization question where to place the reinforced nodes, we focus on $p\leq p_c$ regime.

In addition, we find that $\rho_x^*$ approaches to a constant value when we increase the value of the average intra-degree $z$ for both $p=p_c$ and $p=0.9\cdot p_c$ (see Fig. \ref{fig:best_rhox_k}). Therefore, we conclude that although $\rho^{*}_x$ depends on $r$ and $p$, for high average intra-degree it does not depend on the degree.

\section{Summary}

In summary, we have developed a general percolation framework for studying a new realistic network model of $m$ ER modules.
We have derived the effect of reinforced nodes on the size of the functional component (FC) of our modular network, i.e., the effect of such a decentralization approach on the network's robustness.
Previously, the concept of reinforced nodes has been studied only for a non-modular (single community) network and when placing the reinforced nodes at random while here we addressed for the first time an optimization problem of modular networks and non-random locations.
We find the fraction of reinforced nodes within the inter-connected nodes which provides the largest FC, $\rho_x^*$, by simulations and theory.
We also showed that for a broad range of parameters the value of $\rho_x^*$ is a non-trivial intermediate value (especially near criticality $p_c$) and becomes constant for high average intra-degrees. 
These results may have significant practical applications. For example, they can be used to determine the optimal way to distribute the power generators, in a given electricity infrastructure network (which usually has a modular structure).

 \section*{ACKNOWLEDGMENTS}

We thank the Israel Science Foundation, the Binational Israel-China Science Foundation (Grant No. 3132/19), the BIU Center for Research in Applied Cryptography and Cyber Security, NSF-BSF (Grant No. 2019740), the EU H2020 project RISE (Project No. 821115), the EU H2020 DIT4TRAM, and DTRA (Grant No. HDTRA-1-19-1-0016) for financial support.
D.V. thanks the PBC of the Council for Higher Education of Israel for the Fellowship Grant.

\FloatBarrier

\section{Appendix}

\subsection{Additional figures for Figs. \ref{fig:S vs rho_x at pc}-\ref{fig:best_rhox}}

\begin{figure}[ht!]
\centering
\includegraphics[width=\linewidth]{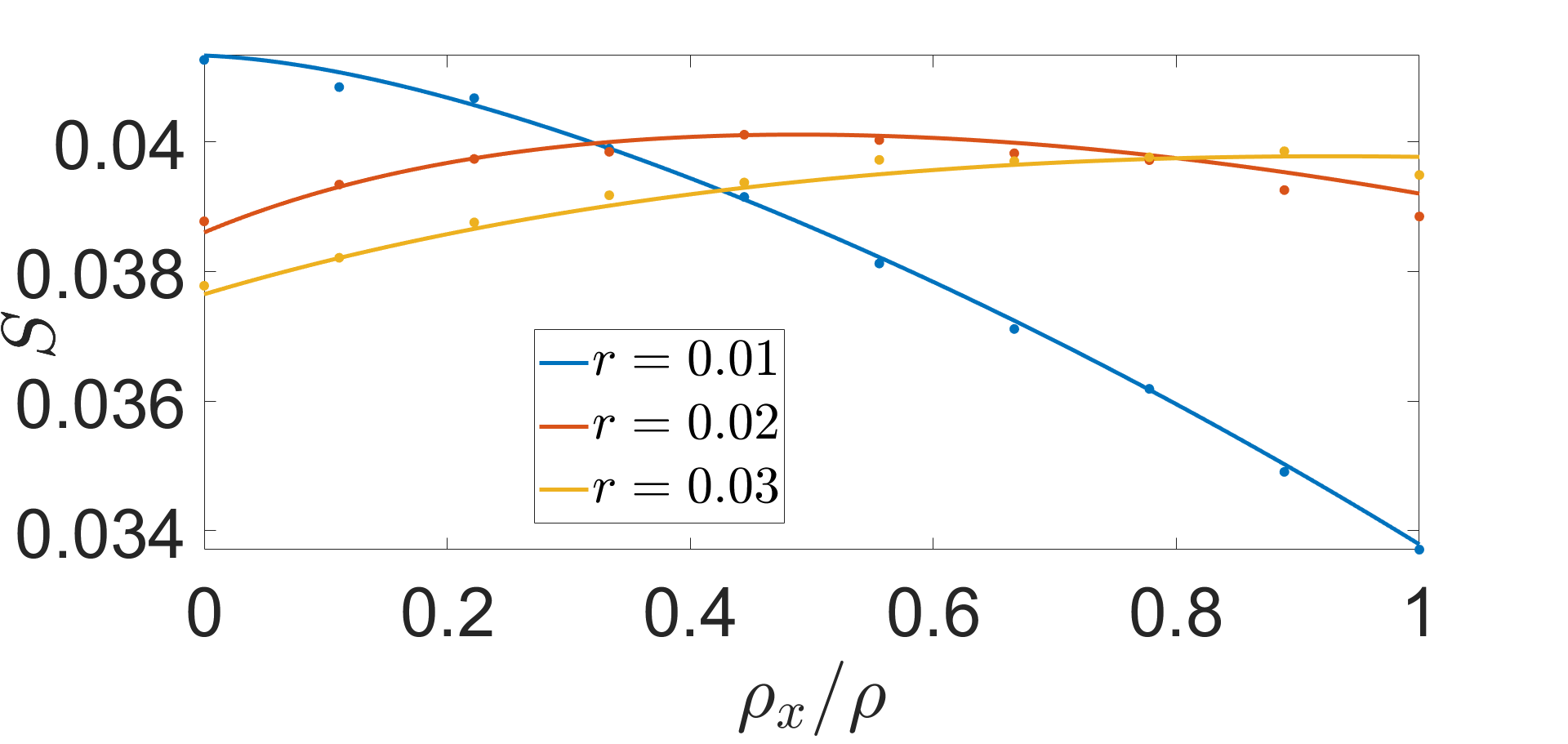}
\caption{The size of the functional component $S$ vs $\rho_x/\rho$, at $p_c$, for several values of $r$. Lines and symbols denote analytical and simulation results, respectively. For simulations and theory, $m=2$, $N_1 = N_2 = 10^6$, $M_{inter} = 5\cdot 10^4$, $z=4$, $p_c=0.25$ and $\rho = 0.01$.}
	\label{fig:S vs rho_x at pc z4}
\end{figure}

In Fig. \ref{fig:S vs rho_x at pc z4} we show that $S$ as a function of $\rho_x/\rho$ for the average intra-degree $z=4$ behaves similarly to the behaviour of $S$ as a function of $\rho_x/\rho$ for $z=3$ (shown in Fig. \ref{fig:S vs rho_x at pc}). 
The figure shows that there is a value $\rho_x^*$ for which the modular network has optimal robustness to random failures. 

\begin{figure}[ht!]
\centering
	\subfloat[]{\includegraphics[width=\linewidth]{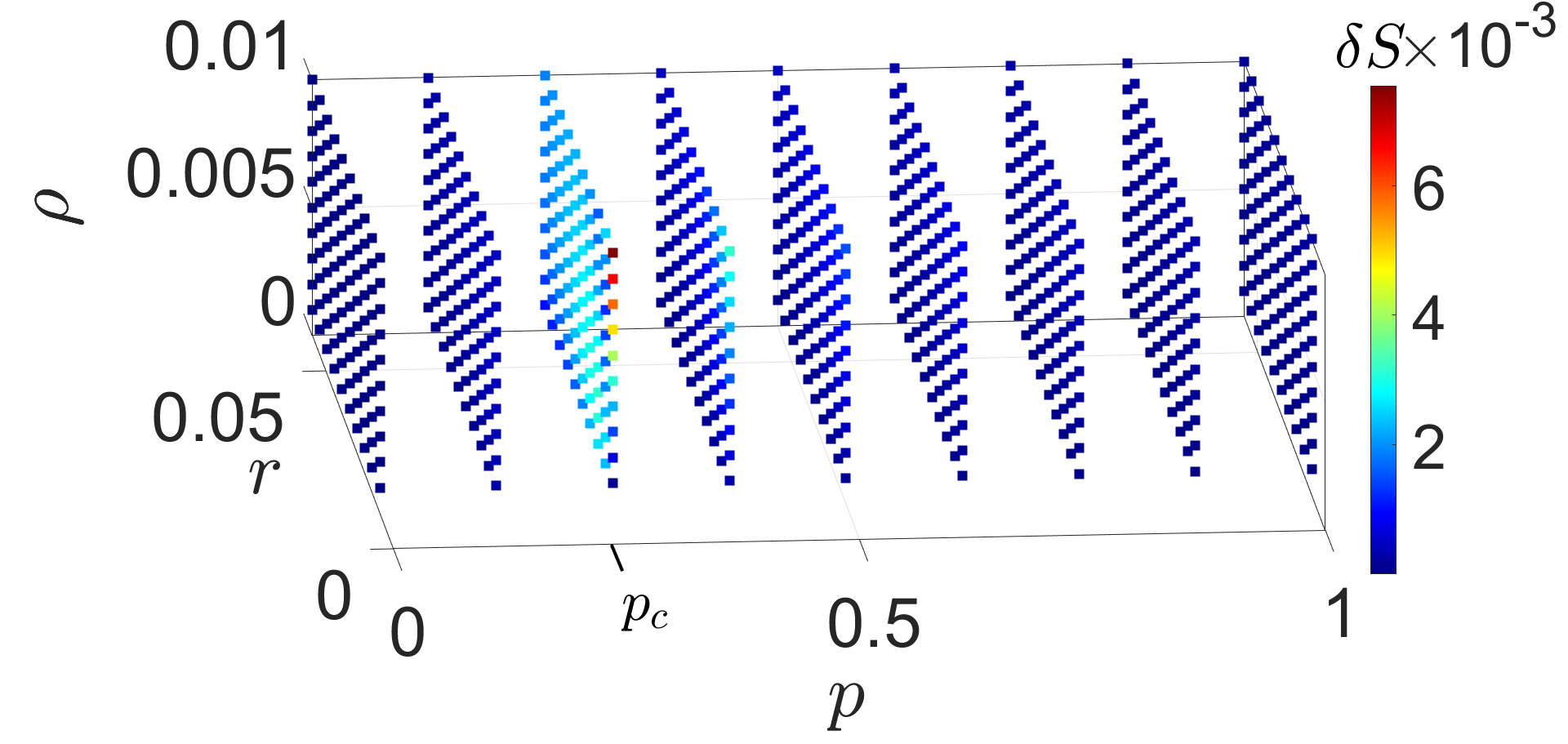}}\\
	\subfloat[]{\includegraphics[width=\linewidth]{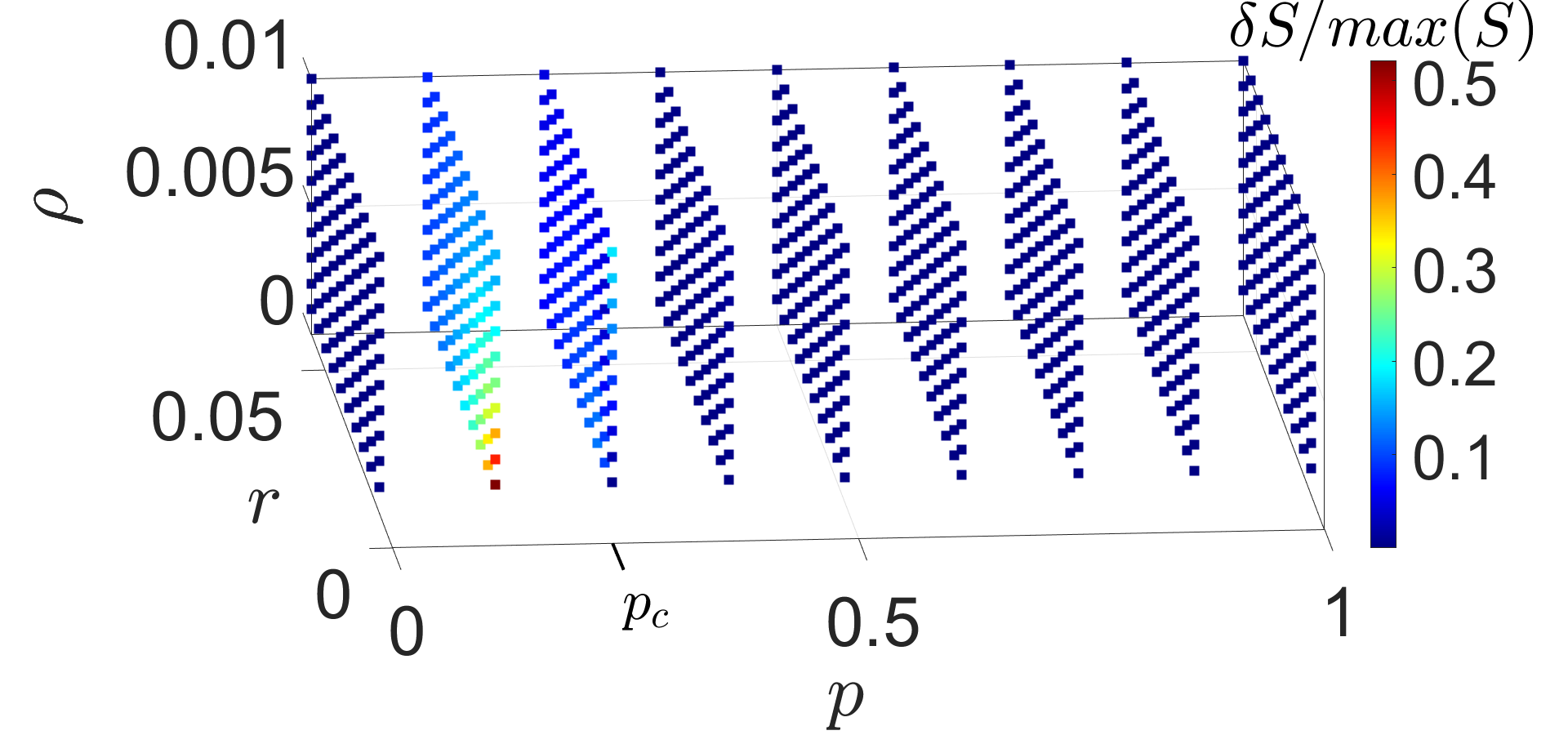}}
\caption{	{\bf (a)} The difference, $\delta S$, between the largest and the smallest values of the functional component size, $S$, that obtained from different values of $\rho_x$. Near $p_c$ the $\delta S$ is maximally.
	{\bf (b)} The values of $\delta S$ from (a) divided by the maximal size of the functional component. Below $p_c$ the $\delta S / max(S)$ is significantly larger than above it.
	For these runs, $m=2$, $N_1 = N_2 = 10^6$, $M_{inter} = 5\cdot 10^4$ and $z=4$.}
	\label{fig:delta S for r_rho_p}
\end{figure}

For understanding in which cases the different locations of the reinforced nodes within the network are significant, we calculate the differences in the size of the functional component between different partitions of the reinforced nodes as shown in Fig. \ref{fig:delta S for r_rho_p}. We obtain that mostly around $p_c$ the location of the reinforced nodes is important. We also obtain the same conclusion when we evaluate the ratio between the differences in the size of the functional component and the maximal functional component value.
It is seen that reinforced nodes can save a large fraction of the network.

\subsection{Results for $m>2$ communities}

\begin{figure}[ht!]
\centering
	\subfloat[]{\includegraphics[width=\linewidth]{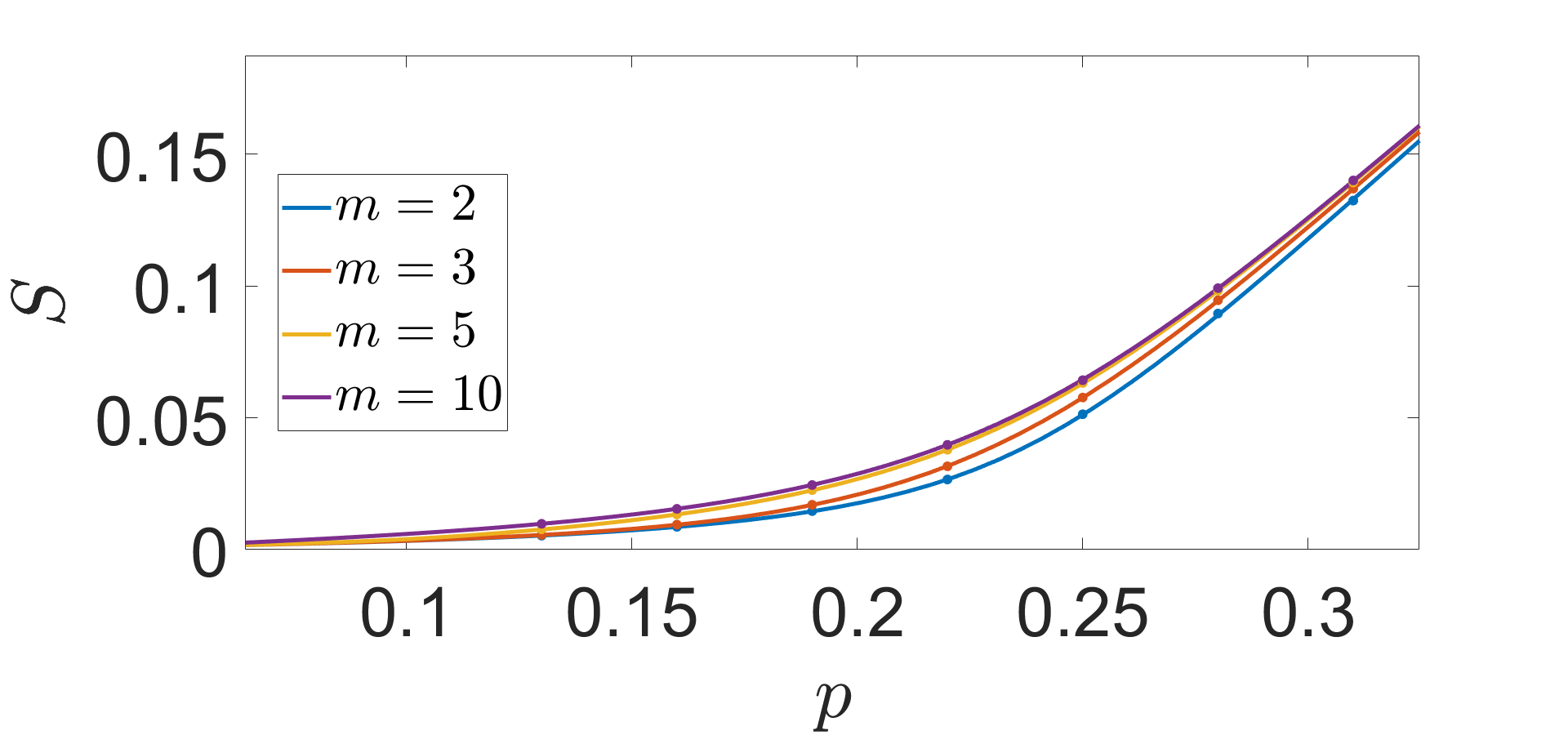}}\\
	\subfloat[]{\includegraphics[width=\linewidth]{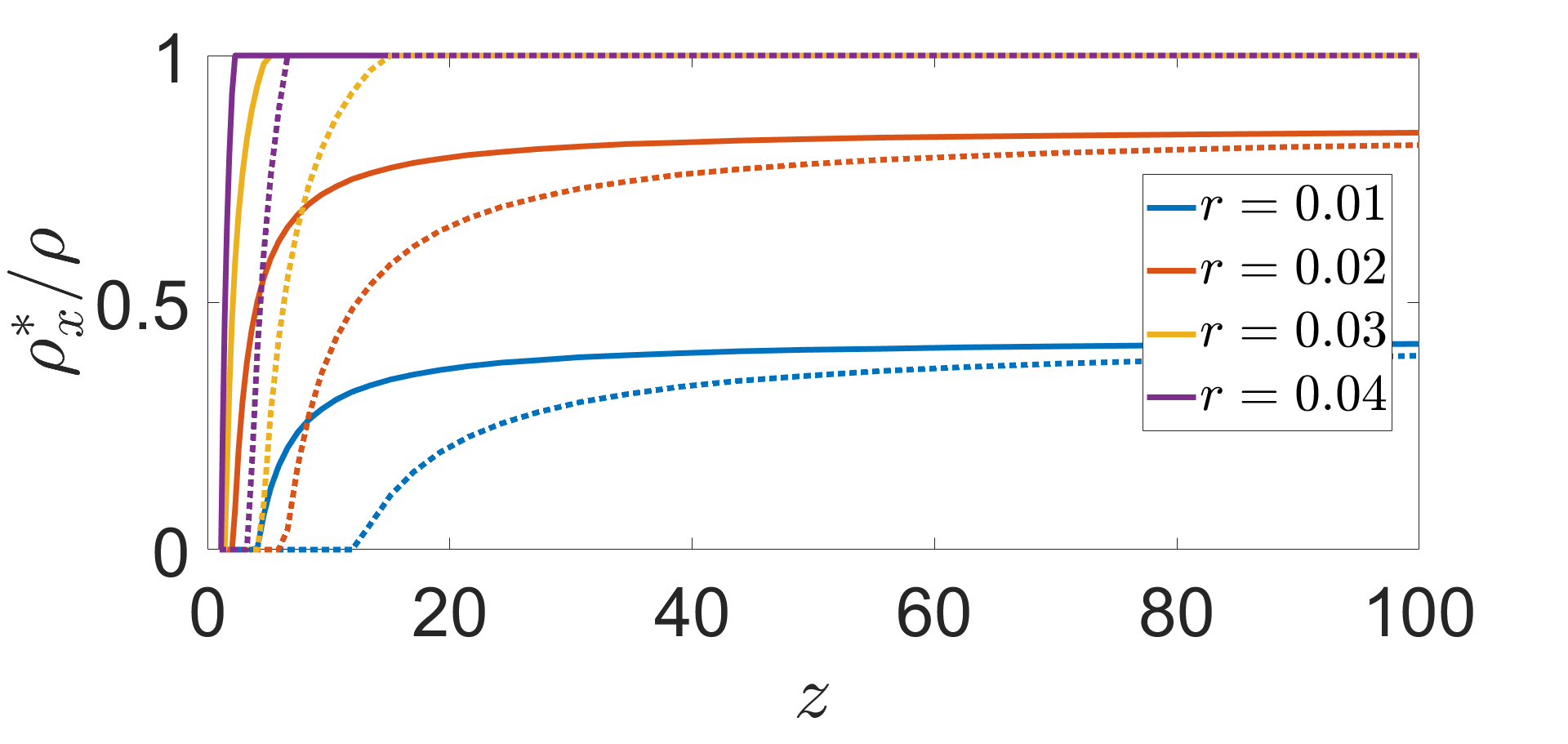}}
\caption{	{\bf (a)} The size of the functional component in an ER modular network as a function of $p$ for different number of communities, $m$, where the reinforced nodes are distributed randomly. Here $z=4$, $r=0.02$ and $\rho=0.02$.
	{\bf (b)} The fraction $\rho^{*}_x$/$\rho$ vs $z$ at the $P_c= 1/z$ for different values of $r$. Here $\rho = 0.01$. The symbols are for $m=4$ communities while the full lines are for $m=2$ communities. The optimal ratio of the reinforced nodes remains constant for high intra-degrees.
	For these runs, each community contains $10^6$ nodes and $M_{inter} = 5\cdot 10^4$.}
	\label{fig:different communities}
\end{figure}

In our research we also tested Eq. (\ref{eq:communities_reinforced}) via simulations for networks with a different number of modules, $m$, containing randomly distributed reinforced nodes. As shown in Fig. \ref{fig:different communities}(a), the results (dots) we obtain are in a very good agreement with the analytical solution (lines) of Eq. (\ref{eq:communities_reinforced}).

Furthermore, we study the optimization conditions for the system with different numbers of modules by using the analytical derived in Eq. (\ref{eq:m_communities_rho_x_rho_o}).
We also found that the value of $\rho_x^*$ , for different number of modules, behaves similarly to Fig. \ref{fig:best_rhox_k}, i.e., also for $m=4$ for higher intra-degrees the value $\rho_x^*$ becomes constant (see Fig. \ref{fig:different communities}(b)).

\FloatBarrier
\bibliographystyle{naturemag_4etal}
\bibliography{references}

\end{document}